\documentclass[prl,aps,floatfix,twocolumn,superscriptaddress,preprintnumbers]{revtex4-1}

\pdfoutput=1
\usepackage{mathrsfs}
\usepackage{amsfonts}
\usepackage{amsmath,amssymb,bm,bbm}
\usepackage[colorlinks=true,linkcolor=blue,citecolor=blue,urlcolor=blue]{hyperref}
\usepackage{epsfig}
\usepackage{graphicx}
\usepackage{slashed}
\usepackage{booktabs}
\usepackage{tabu}
\usepackage[dvipsnames]{xcolor}
\usepackage[normalem]{ulem}
\usepackage{tikz}
\usepackage{soul}
\usetikzlibrary{shapes,arrows}
\usetikzlibrary{positioning}
\usepackage{ulem}
\usepackage{hhline}
\usepackage{sidecap}
\sidecaptionvpos{figure}{c}
\sidecaptionvpos{table}{c}

\newcommand{\diff}[1]{\mathrm{d}#1}

\newcommand{\sa}{\sphericalangle}

\newcommand{\chired}{{\chi^2_{\rm red}}}

\allowdisplaybreaks

\begin{document}

\preprint{JLAB-THY-23-3859}
\title{Transversity distributions and tensor charges of the nucleon: \\ extraction from dihadron production and their universal nature}

\newcommand*{\TU}{Department of Physics, SERC, Temple University, Philadelphia, Pennsylvania 19122, USA}\affiliation{\TU}
\newcommand*{\LVC}{Department of Physics, Lebanon Valley College, Annville, Pennsylvania 17003, USA}\affiliation{\LVC}
\newcommand*{\PSU}{Division of Science, Penn State University Berks, Reading, Pennsylvania 19610, USA}\affiliation{\PSU}
\newcommand*{\JLAB}{Jefferson Lab, Newport News, Virginia 23606, USA}\affiliation{\JLAB}
\newcommand*{\RIKEN}{RIKEN BNL Research Center, Upton, New York 11973, USA}
\affiliation{\RIKEN}

\author{C.~Cocuzza}\affiliation{\TU}
\author{A.~Metz}\affiliation{\TU}
\author{D.~Pitonyak}\affiliation{\LVC}
\author{A.~Prokudin}\affiliation{\PSU}\affiliation{\JLAB}
\author{N.~Sato}\affiliation{\JLAB}
\author{R.~Seidl}\affiliation{\RIKEN}
\collaboration{{\bf Jefferson Lab Angular Momentum (JAM) Collaboration}}

\begin{abstract}
We perform the first global quantum chromodynamics (QCD) analysis of dihadron production for a comprehensive set of data in electron-positron annihilation, semi-inclusive deep-inelastic scattering, and proton-proton collisions, from which we extract simultaneously the transversity distributions of the nucleon and $\pi^+\pi^-$ dihadron fragmentation functions. 
We incorporate in our fits known theoretical constraints on transversity, namely, its small-$x$ asymptotic behavior and the Soffer bound.
We furthermore show that lattice-QCD results for the tensor charges can be successfully included in the analysis.  This resolves the previously reported incompatibility between the tensor charges extracted from dihadron production data and lattice QCD.
We also find agreement with results for the transversity and tensor charges obtained from measurements on single-hadron production.
Overall, our work demonstrates for the first time the universal nature of all available information for the transversity distributions and the tensor charges of the nucleon.
\end{abstract}

\date{\today}
\maketitle

{\it Introduction} ---\
%
Protons and neutrons (nucleons) have several fundamental charges that govern basic nuclear processes.  For example, the beta decay rate of a free neutron depends on the ratio of axial-vector and vector charges.  The ability to measure certain beyond the Standard Model (BSM) couplings in this reaction (see, e.g., Refs.~\cite{Herczeg:2001vk,Erler:2004cx,Severijns:2006dr,Cirigliano:2013xha,Courtoy:2015haa,Gonzalez-Alonso:2018omy})
relies on knowledge of another fundamental charge of the nucleon:~the isovector tensor charge $g_T$, which is given by the up and down quark tensor charges, $\delta u$ and $\delta d$, through $g_T = \delta u - \delta d$.
The quark tensor charges themselves are ingredients for BSM physics in computing the nucleon electric dipole moment from those of the quarks (see, e.g., Refs.~\cite{Erler:2004cx,Pospelov:2005pr,Yamanaka:2017mef,Liu:2017olr}).  Therefore, it is crucial to have precise values for $\delta u$, $\delta d$, and $g_T$ and compatibility between different techniques used for their determination.

The main approaches to obtaining the tensor charges within QCD are phenomenological analyses of experimental data~\cite{Anselmino:2007fs, Anselmino:2008jk, Bacchetta:2012ty, Anselmino:2013vqa, Goldstein:2014aja, Anselmino:2015sxa, Radici:2015mwa, Kang:2015msa, Lin:2017stx, Radici:2018iag, Benel:2019mcq, DAlesio:2020vtw, Cammarota:2020qcw, Gamberg:2022kdb}, ab initio calculations in lattice QCD (LQCD)~\cite{Gupta:2018qil, Gupta:2018lvp, Yamanaka:2018uud, Hasan:2019noy, Alexandrou:2019brg, Harris:2019bih, Horkel:2020hpi, Alexandrou:2021oih, Park:2021ypf, Tsuji:2022ric, Bali:2023sdi, Smail:2023eyk}, and model calculations~\cite{He:1994gz, Barone:1996un, Schweitzer:2001sr, Gamberg:2001qc, Pasquini:2005dk, Wakamatsu:2007nc, Lorce:2007fa, Yamanaka:2013zoa, Pitschmann:2014jxa, Xu:2015kta, Wang:2018kto, Liu:2019wzj}.
For the former, parton distribution functions (PDFs) are essential, as they describe the one-dimensional momentum-space structure of nucleons in terms of the longitudinal momentum fraction $x$ of the parton (quark or gluon). At leading twist, the nucleon is characterized by the spin-averaged PDF $f_1$, the helicity PDF $g_1$, and the transversity PDF $h_1$.
The transversity PDF~\cite{Ralston:1979ys} is of particular interest here due to its relation to the tensor charges:
\begin{align}
\delta u = \int_0^1 \!\!\diff x ~h_1^{u_v}(x;\mu)\,, ~~~~ 
\delta d = \int_0^1 \!\!\diff x ~h_1^{d_v}(x;\mu)\,,
\label{e.tensorcharge}
\end{align}
where $h_1^{q_v} \equiv h_1^q - h_1^{\bar{q}}$ are the valence distributions, and $\mu$ is the renormalization scale.

The transversity PDF quantifies the degree of transverse polarization of quarks within a transversely polarized nucleon. The chiral-odd nature of transversity makes it difficult to extract (relative to the spin-averaged and helicity PDFs) since it must couple to another chiral-odd function.
This can be achieved through single-hadron production, either by exploiting transverse momentum dependent (TMD) factorization using, e.g., the Collins effect~\cite{Collins:1992kk, Bacchetta:2006tn} in semi-inclusive deep-inelastic scattering (SIDIS)~\cite{Anselmino:2007fs, Anselmino:2008jk, Anselmino:2013vqa, Anselmino:2015sxa, Kang:2015msa, Lin:2017stx, DAlesio:2020vtw, Cammarota:2020qcw, Gamberg:2022kdb}, or collinear subleading-twist (twist-3) factorization in hadronic collisions~\cite{Metz:2012ct, Kanazawa:2014dca,Gamberg:2017gle, Cammarota:2020qcw, Gamberg:2022kdb}.  
Alternatively, one can consider dihadron production in leading-twist collinear factorization, where the transversity PDF couples to dihadron fragmentation functions (DiFFs)~\cite{Collins:1993kq,Bianconi:1999cd,Bacchetta:2002ux,Boer:2003ya,Bacchetta:2003vn,Bacchetta:2004it,Courtoy:2012ry,Bacchetta:2012ty,Radici:2015mwa,Radici:2018iag,Matevosyan:2018icf,Benel:2019mcq}, in particular to the so-called interference fragmentation function (IFF) $H_1^{\sa}$.
The DiFF $D_1$ is also a necessary part of the relevant experimental observables~\footnote{Generally the DiFFs are written with a superscript $h_1 h_2$ to denote the hadron pair.  Here we consider only $\pi^+ \pi^-$ DiFFs and drop this superscript to simplify the notation.}.
There is noted tension between the results for $\delta u$ and $g_T$ from the DiFF approach~\cite{Radici:2018iag,Benel:2019mcq} and recent extractions using the TMD/collinear twist-3 method (JAM3D)~\cite{Cammarota:2020qcw, Gamberg:2022kdb}, as well as  the tensor charges computed in LQCD.

The resolution of this issue is of utmost importance.  Towards this goal, we perform the first simultaneous global QCD analysis (JAMDiFF) of the $\pi^+ \pi^-$ DiFFs and transversity PDFs from electron-positron ($e^+ e^-$) annihilation, SIDIS, and proton-proton ($pp$) data.
We include, for the first time, the Belle cross section~\cite{Belle:2017rwm}, the latest measurements from STAR~\cite{STAR:2017wsi}, and all kinematic variable binnings for the relevant processes under consideration, making this the most comprehensive study of dihadron observables to date.
We implement theoretical constraints at small $x$~\cite{Kovchegov:2018zeq} and large~$x$~\cite{Soffer:1994ww} where experimental data is absent.
This allows us to meaningfully calculate the integrals in Eq.~(\ref{e.tensorcharge}) and include LQCD data into the fit, similar to what has been done in Refs.~\cite{Lin:2017stx, Gamberg:2022kdb}. We then examine the compatibility between phenomenological results based on the dihadron approach, those in the TMD/collinear twist-3 framework, and LQCD computations.  
A companion paper providing further details on certain aspects of this analysis can be found in Ref.~\cite{Cocuzza:2023vqs}.

{\it Theoretical Framework ---}\ 
%
Our theoretical framework is based on leading-twist collinear factorization, at leading-order (LO) in the strong coupling constant $\alpha_s$, for high-energy $\pi^+ \pi^-$ production from $e^+ e^-$ annihilation, SIDIS, and $pp$ collisions.
The DiFFs under consideration depend upon the fractional momentum of the dihadron pair $z$ and its invariant mass $M_h$~\footnote{DiFFs at the fully unintegrated level depend on several more variables besides $(z,M_h)$.  Since these more differential objects do not enter our analysis, we will refer to the $(z,M_h)$-dependent functions (which are also sometimes called ``extended'' DiFFs) simply as DiFFs without any ambiguity.}. 
We note that the formulas in this section use a new definition of the DiFFs that has a number density interpretation~\cite{Pitonyak:2023gjx, Cocuzza:2023vqs}.  Such a change does not affect our ability to make quantitative comparisons of the extracted transversity PDFs to those of other groups (see Figs.~\ref{f.tpdfs}, \ref{f.tensorcharge} below).

We begin with discussing the processes
$e^+ e^- \to (\pi^+ \pi^-)\, X$ and $e^+ e^- \to (\pi^+ \pi^-) (\pi^+ \pi^-) \,X$ at center-of-mass (COM) energy $\sqrt{s}$, which sets the hard (renormalization) scale for these reactions,
where $X$ represents all undetected particles.
For the first reaction, the cross section is given by~\cite{Pitonyak:2023gjx} $\diff \sigma/ \diff z \,\diff M_h = 4\pi N_c \alpha_{\rm em}^2\sum_q \bar{e}_q^2\, D_1^{q}(z,M_h)/(3s)$,
where the sum is over quarks and antiquarks, $\alpha_{\rm em}$ is the fine structure constant, $N_c=3$ is the number of quark colors, and $\bar{e}_q^2 = e_q^2 + \ldots\,$, where $e_q$ is the charge of the quark flavor $q$ in units of the elementary charge, and the ellipsis represents the contributions from $\gamma Z$ and $ZZ$ exchanges.
For the process with two sets of hadron pairs  detected in opposite hemispheres,  one can measure an azimuthal correlation, known as the Artru-Collins asymmetry \cite{Artru:1995zu}, given by~\cite{Boer:2003ya, Matevosyan:2018icf}
\begin{align}
A^{e^+ e^-} =\frac{ \sin^2 \theta \sum_q e_q^2 \, H_1^{\sa,q}(z,M_h) H_1^{\sa,\bar{q}}(\bar{z},\overline{M}_h)}{(1+\cos^2\theta)\sum_q e_q^2 \, D_1^{q}(z,M_h) D_1^{\bar{q}}(\bar{z},\overline{M}_h)}\,,
\label{e.a12R}
\end{align}
where $\overline{M}_h$ and $\bar{z}$ are the invariant mass and fractional energy of the second dihadron, and $\theta$ is defined as the polar angle between the beam axis and the reference axis in the COM system.

In the SIDIS process $\ell \,N^{\uparrow} \to \ell'\, (\pi^+ \pi^-)\, X$, a lepton with 4-momentum $k$ scatters off a transversely polarized proton ($N=p$) or deuteron ($N=D$) with 4-momentum $P$ and is detected with 4-momentum $k'$ along with a $\pi^+ \pi^-$ pair.
The intermediate photon has virtuality $Q^2 = -q^2 = -(k -k')^2$, and the inelasticity is $y \equiv P \cdot q/P \cdot k$.
The asymmetry can be written as \cite{Bacchetta:2002ux, Bacchetta:2003vn}
\begin{align}
A_{UT}^{\rm SIDIS} = c(y) \frac{\sum_q e_q^2\,h_1^q(x)\,H_1^{\sphericalangle,q}(z,M_h)}{\sum_q e_q^2\,f_1^q(x)\,D_1^{q}(z,M_h)}\,.
\label{e.AUTSIDIS}
\end{align}
The factor $c(y)$ depends on the definition of the asymmetry:~it is $1$ for COMPASS \cite{COMPASS:2023cgk} and $-(1-y)/(1-y+\frac{y^2}2)$ for HERMES \cite{HERMES:2008mcr}.
We set the hard scale to be $\mu = Q$, and $f_1$ is taken from Ref.~\cite{Cocuzza:2022jye}.
For the deuteron target we neglect nuclear corrections and take the PDFs in Eq.~(\ref{e.AUTSIDIS}) to be the average of proton and neutron PDFs, using isospin symmetry to relate the neutron PDFs to those of the proton.

We also consider the process $p^{\uparrow}p \to (\pi^+ \pi^-)\, X$, where a transversely polarized proton with 4-momentum $P_A$ collides with an unpolarized proton with 4-momentum $P_B$ producing an outgoing dihadron pair with 4-momentum $P_h$ (with $P_{hT}$ the magnitude of the transverse component) and pseudo-rapidity $\eta$.
The asymmetry reads
\begin{equation}
A_{UT}^{pp} = \frac{\mathcal{H}(M_h,P_{hT},\eta)}{\mathcal{D}(M_h,P_{hT},\eta)}\,,
\end{equation}
where one has \cite{Bacchetta:2004it}
\begin{align}
\mathcal{H}(M_h,P_{hT},\eta) &= 2 P_{hT} \sum_i \sum_{a,b,c,d}\int_{x_{a}^{\rm min}}^1 \! \diff x_a\int_{x_{b}^{\rm min}}^1 \! \frac{ \diff x_b}{z} \label{e.ppH} \\
&\!\times h_1^a(x_a)\,f_1^b(x_b)\frac{\diff \Delta\hat{\sigma}_{a^\uparrow b\to c^\uparrow d}}{\diff \hat{t}}\,H_1^{\sa,c}(z,M_h)\,, \notag
\end{align}
with $x_a$ ($x_b$) the momentum fraction of proton $A$ ($B$) carried by the parton. 
The sum $\sum_i$ is over all partonic sub-processes, with $\sum_{a,b,c,d}$ summing over all possible parton flavor combinations for a given channel.  The relevant Mandelstam variables are $s = (P_A+P_B)^2$, $t = (P_A-P_h)^2$, and $u = (P_B-P_h)^2$.  The total momentum fraction of the dihadron is fixed by $z=\frac{P_{hT}}{\sqrt{s}}(\frac{x_ae^{-\eta}+x_be^\eta}{x_a x_b})$.
The expression for $\mathcal{D}(M_h,P_{hT},\eta)$ is the same as Eq.~(\ref{e.ppH}) with the replacements $h_1^a \to f_1^a$, $\diff \Delta\hat{\sigma}_{a^\uparrow b\to c^\uparrow d} \to \diff \hat{\sigma}_{ab\to cd}$, and $H_1^{\sa,c} \to D_1^c$.
The hard scale is taken to be $\mu = P_{hT}$.
The hard (perturbative) partonic cross sections $\diff \hat{\sigma}$ and $\diff \Delta \hat{\sigma}$~\cite{Bacchetta:2004it} depend on $\hat{s}=x_ax_b s$, $\hat{t} = x_a s/z$, and $\hat{u} = x_b u/z$.
The limits on the integration are
\vspace{-0.1cm}
\begin{align}
x_{a}^{\rm min} = \frac{P_{hT} e^\eta}{\sqrt{s}-P_{hT}e^{-\eta}}\,, ~~ x_{b}^{\rm min} = \frac{x_a P_{hT} e^{-\eta}}{x_a\sqrt{s}-P_{hT} e^\eta}\,.
\end{align}

{\it Methodology ---}\ 
%
Our extraction is based on Bayesian inference, with the posterior distribution $\mathcal{P}
\propto \mathcal{L}\,\cdot \,\pi$ (where $\mathcal{L}=\exp(-\chi^2/2)$ is the likelihood function and $\pi$ is the prior  -- see Ref.~\cite{Cocuzza:2023vqs} for details), using the Monte Carlo techniques developed in previous JAM analyses~\cite{Sato:2016tuz, Sato:2016wqj, Ethier:2017zbq, Sato:2019yez,Cammarota:2020qcw, Moffat:2021dji, Cocuzza:2021cbi, Cocuzza:2021rfn, Cocuzza:2022jye,Gamberg:2022kdb}.
We choose to parameterize the PDFs $h_1^{u_v}$, $h_1^{d_v}$, and $h_1^{\bar{u}} = -h_1^{\bar{d}}$ at the input scale $\mu_0 = 1 \, \textrm{GeV}$ using the template function
\begin{eqnarray}
T(x,\mu_0) = \frac{N x^\alpha (1-x)^\beta (1 + \gamma \sqrt{x} + \delta x)}{\int_0^1 \diff x ~ x^\alpha (1-x)^\beta (1 + \gamma \sqrt{x} + \delta x)}\,,
\label{e.template}
\end{eqnarray}
where $N$, $\alpha$, $\beta$, $\gamma$, and $\delta$ are fit parameters.
The template is normalized to the first moment in order to largely decorrelate the normalization and shape parameters.
The relation between the antiquarks is based on large-$N_c$ predictions~\cite{Pobylitsa:2003ty}. 
We utilize it since there are only three unique observables to constrain the transversity PDFs (SIDIS on proton and deuteron and $pp$ collisions), restricting us to extracting three independent functions. (Hypothetically, evolution could disentangle more flavors, but that is not feasible here due to the relatively slow evolution of the transversity PDFs and the large errors and restricted kinematic coverage of the experimental data.)
In contrast to the previous DiFF extractions in Refs.~\cite{Courtoy:2012ry,Radici:2015mwa} that used a considerably more complicated functional form, we use the template Eq.~(\ref{e.template}), with $x \to z$, to parameterize the $z$ dependence of $D_1(z,M_h)$ and $H_1^{\sa}(z,M_h)$.  This is repeated on a grid of $M_h$ and interpolated to obtain the DiFFs at any value of $M_h$.
We evolve the transversity PDFs~\cite{Baldracchini:1980uq, Artru:1989zv, Blumlein:2001ca, Stratmann:2001pt} and DiFFs/IFF~\cite{Pitonyak:2023gjx} using the DGLAP evolution equations with LO splitting functions. 
Further details about the phenomenological methodology, especially regarding the DiFF analysis, can be found in Ref.~\cite{Cocuzza:2023vqs}. There we also discuss tests of parameterization bias but caution here that with those tests we do not exhaust all possible functional forms one could choose for the transversity PDFs and DiFFs/IFF.

We include theoretical constraints at small~$x$ and large~$x$ where experimental measurements are not available.
We impose the Soffer bound $|h_1^q(x)| \leq \frac{1}{2} ( f_1^q(x) + g_1^q(x) )$ \cite{Soffer:1994ww}, 
with $f_1^q$ and $g_1^q$ taken from Ref.~\cite{Cocuzza:2022jye} (with positivity enforced), which is primarily relevant in the unmeasured $x \gtrsim 0.3$ region.
We enforce the Soffer bound on each Monte Carlo replica by using a Bayesian prior that in effect penalizes the $\chi^2$ function~\cite{Ball:2013lla} when it is violated at the input scale, which is justified since the Soffer bound holds under evolution~\cite{Kamal:1996gqi,Barone:1997fh,Vogelsang:1997ak}.
We also limit the small-$x$ behavior of our parameterization, governed by the $\alpha$ parameter in Eq.~(\ref{e.template}).
Theoretical calculations have placed limits on this parameter as $x \rightarrow 0$ (ignoring saturation effects)~\cite{Kovchegov:2018zeq}:
\begin{align}
\alpha \xrightarrow[]{x \to 0} 1 - 2 \sqrt{\frac{\alpha_s N_c} {2 \pi}}\,.
\label{e.alpha}
\end{align}
Therefore, at the input scale we demand $\alpha = 0.17$ with a 50\% uncertainty (allowing $ 0.085 < \alpha < 0.255$), due to unaccounted for $1/N_c$  and next-to-leading order (NLO) corrections \cite{Kovchegov:2023pc,Kirschner:1996jj}, for both the valence quarks and antiquarks.
Ultimately, we find that the average values for $\alpha$ lie near the center of this range with no strong saturation at the bounds.

The unmeasured regions (small $x$ and large $x$) are important when including LQCD results for the tensor charge, as the experimental data is primarily sensitive to the range $0.005 \lesssim x \lesssim 0.3$.
When including LQCD data into the fits, the aforementioned restrictions at small~$x$ (Eq.~(\ref{e.alpha})) and large~$x$ (the Soffer bound) guarantee reasonable physical behavior of the transversity PDFs.
In the absence of these constraints, the PDFs in the unmeasured regions are subject to extrapolation errors which are entirely dependent upon the choice of parameterization, causing extremely large uncertainties for the tensor charges and making agreement with LQCD trivial. With these constraints, an analysis including both experimental and LQCD data is a stringent test of their agreement.  That is, the conclusion that experimental data and LQCD are compatible is robust under changes in our assumptions at small $x$ and large $x$ given the currently available measurements.

{\it Data and Quality of Fit ---}\
To constrain the DiFFs, we use data from Belle on the $e^+ e^-$ dihadron cross section~\cite{Belle:2017rwm} and Artru-Collins asymmetry~\cite{Belle:2011cur}.
In order to achieve flavor separation for $D_1$, we supplement the measurements of Ref.~\cite{Belle:2017rwm} with data from the event generator PYTHIA \cite{Sjostrand:2001yu} used as a Bayesian prior. We use the tunes studied in Ref.~\cite{Belle:2017rwm} to generate systematic errors as well as different COM energies to constrain $D_1^g$ through scaling violations.  Further details are presented in Ref.~\cite{Cocuzza:2023vqs}.

\begin{table}
\resizebox{0.40\textwidth}{!} {
\begin{tabular}{l | c | c | c }
\hhline{====}
\vspace{-0.3cm} &   \\
& & \multicolumn{2}{c}{$\chired$} \\
\vspace{-0.3cm} &   \\
Experiment &  $N_{\rm dat}$  & w/ LQCD & no LQCD \\
\hline
Belle (cross section) \cite{Belle:2017rwm}     &  1094    & 1.01  & 1.01 \\
Belle (Artru-Collins) \cite{Belle:2011cur}     &  183     & 0.74  & 0.73 \\
\hline
HERMES \cite{HERMES:2008mcr}                   &  12      & 1.13  & 1.10 \\
COMPASS ($p$) \cite{COMPASS:2023cgk}           &  26      & 1.24  & 0.75 \\
COMPASS ($D$) \cite{COMPASS:2023cgk}           &  26      & 0.78  & 0.76 \\ \hline
STAR (2015) \cite{STAR:2015jkc}                &  24      & 1.47  & 1.67 \\
STAR (2018) \cite{STAR:2017wsi}                &  106     & 1.20  & 1.04 \\ \hline
ETMC $\delta u$ \cite{Alexandrou:2019brg}      & 1        & 0.71  & ---  \\
ETMC $\delta d$ \cite{Alexandrou:2019brg}      & 1        & 1.02  & ---  \\
PNDME $\delta u$ \cite{Gupta:2018lvp}          & 1        & 8.68  & ---  \\
PNDME $\delta d$ \cite{Gupta:2018lvp}          & 1        & 0.04  & ---  \\
\hline
\vspace{-0.3cm} & & &  \\
\textbf{Total} $\boldsymbol{\chired}$ ($N_{\rm dat}$)  &  & {\bf 1.01} (1475) & {\bf 0.98} (1471) \\
\hhline{====}
\end{tabular}}
\caption
{Summary of $\chired$ values for the fits with and without LQCD data.
}
\label{t.chi2}
\end{table}

The transversity PDFs are constrained by SIDIS and $pp$ data. The SIDIS data is from HERMES~\cite{HERMES:2008mcr} and COMPASS~\cite{COMPASS:2023cgk}, and we use all three binnings ($x$, $z$, $M_h$). 
The $pp$ data is from STAR, at both $\sqrt{s} = 200 \, \textrm{GeV}$~\cite{STAR:2015jkc} and $\sqrt{s} = 500 \, \textrm{GeV}$~\cite{STAR:2017wsi}.  
The $\sqrt{s} = 200 \, \textrm{GeV}$ data is provided with three different upper cuts (0.2, 0.3, 0.4) on the opening angle $R$ of the pion pair, with 0.3 treated as the default.  
This cut is used to filter out pion pairs that do not originate from a single parton.
We use the data corresponding to $R < 0.3$ and have verified that the changes to the extracted functions by instead using data from the other cuts are negligible compared to the statistical uncertainties of the functions themselves.
The $\sqrt{s} = 500 \, \textrm{GeV}$ data have a larger opening angle cut of $R < 0.7$.  However, the increased energy means that gluon radiation is occurring at wider angles, allowing the dihadron pair to still be considered as originating from a single parton even with a larger $R$-cut value.  
The $pp$ data is binned in $P_{hT}$, $M_h$, and $\eta$, with the results (often) provided for both $\eta > 0$ and $\eta < 0$ when binned in $P_{hT}$ or $M_h$, and we include all binnings.

We also consider the inclusion of LQCD data as a Bayesian prior in the analysis.  We restrict ourselves to results at the physical pion mass with $2 \!+\! 1 \!+\! 1$ flavors, where calculations are available from ETMC \cite{Alexandrou:2019brg} and PNDME \cite{Gupta:2018lvp} on $\delta u$, $\delta d$, and $g_T$.
We choose to include $\delta u$ and $\delta d$ rather than $g_T$ in order to provide flavor separation.
Below we will discuss the results without and with these LQCD calculations, referred to as JAMDiFF (no LQCD) and JAMDiFF (w/ LQCD), respectively.

The reduced $\chi^2$ ($\chired$), calculated using the mean theory value, for the two scenarios are shown in Table \ref{t.chi2}.
We re-emphasize that we have performed a simultaneous global analysis of DiFFs and transversity PDFs, where, unlike previous work~\cite{Bacchetta:2012ty, Radici:2015mwa, Radici:2018iag, Benel:2019mcq}, the parameters for the DiFFs are not fixed (from a fit of only $e^+ e^-$ annihilation) but allowed to be free along with the transversity PDF parameters.
We have also studied an exhaustive set of available data on dihadron observables, which includes, for the first time, the Belle cross section~\cite{Belle:2017rwm}, the $\sqrt{s} = 500 \, \textrm{GeV}$ measurements from STAR~\cite{STAR:2017wsi}, and all kinematic variable binnings for the relevant processes under consideration, amassing 1471 experimental data points.
Both with and without LQCD we are able to describe all of the experimental data very well.
We will discuss the $\chired$ for the LQCD fit below in conjunction with the tensor charge results.

{\it Transversity PDFs ---}\
%
\begin{figure}[t]
\includegraphics[width=0.48\textwidth]{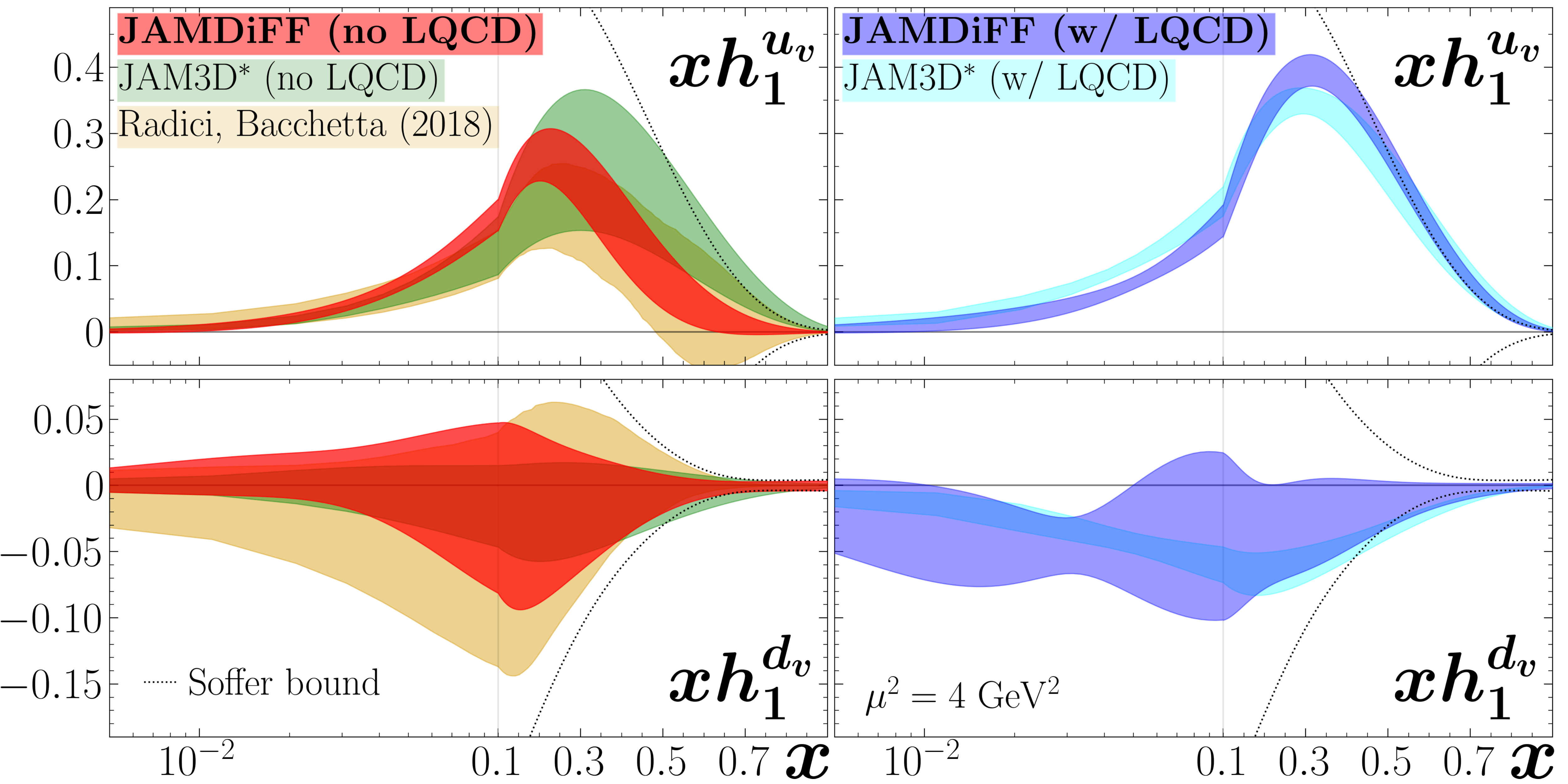}
\caption
{Transversity PDFs $x h_1^{u_v}$ (top row) and $x h_1^{d_v}$ (bottom row) plotted as a function of $x$ at the scale $\mu^2 = 4$ GeV$^2$.
Our results (JAMDiFF) are shown at 1$\sigma$ both without (red) and with (blue) LQCD included in the fit and are compared to those from JAM3D$^*$ \cite{Gamberg:2022kdb,Note3} at 1$\sigma$ without (green) and with (cyan) LQCD as well as RB18 \cite{Radici:2018iag} (gold, 90\% confidence level).
The Soffer bound is indicated by the dashed black lines.}
\label{f.tpdfs}
\end{figure}
The following results for the no LQCD and w/ LQCD fits are produced from over 900 replicas each.
In Fig.~\ref{f.tpdfs} we compare our results with and without LQCD for the transversity valence distributions to those from Radici, Bacchetta \cite{Radici:2018iag} (RB18) (whose analysis did not consider the inclusion of lattice data) and a version of JAM3D that has been slightly updated from Ref.~\cite{Gamberg:2022kdb} (see the footnote~\footnote{In order to align with the methodology of JAMDiFF, we show here results from the JAM3D analysis that are slightly updated from Ref.~\cite{Gamberg:2022kdb}:~antiquark transversity PDFs are now included (with $h_1^{\bar{d}} = - h_1^{\bar{u}}$), the small-$x$ constraint Eq.~(\ref{e.alpha}) is imposed, and, for the fit with LQCD, $\delta u$ and $\delta d$ from ETMC and PNDME are used (instead of only the $g_T$ data point from ETMC).}) that we will refer to as JAM3D$^*$.
For the no LQCD results we agree with RB18 within errors, but with a larger $h_1^{u_v}$ in the region $ 0.04 \lesssim x \lesssim 0.3$.

Comparing to JAM3D$^*$ without LQCD, we find that our distributions agree, except $h_1^{u_v}$ from JAM3D$^*$ has a preference to be slightly larger at higher $x$.
When including LQCD, the results for $h_1^{d_v}$ remain in agreement, while our result for $h_1^{u_v}$ is slightly larger than JAM3D$^*$ in the $x \gtrsim 0.3$ valence region and slightly smaller for $0.01 \lesssim x \lesssim 0.1$.
While the inclusion of the LQCD data fixes the moments of the valence transversity PDFs, it is non-trivial to find that the $x$-dependence of the JAMDiFF and JAM3D$^*$ distributions also largely match. (A comparison with LQCD results for the $x$-dependence of  transversity~\cite{Alexandrou:2019lfo, HadStruc:2021qdf} can be found in Ref.~\cite{Gamberg:2022kdb}.)  Our extracted transversity PDFs (and DiFFs) can be found in a {\tt github} library~\cite{github-repo} and a {\tt google colab} notebook~\cite{google-colab}.

\begin{figure}[t]
\includegraphics[width=0.48\textwidth]{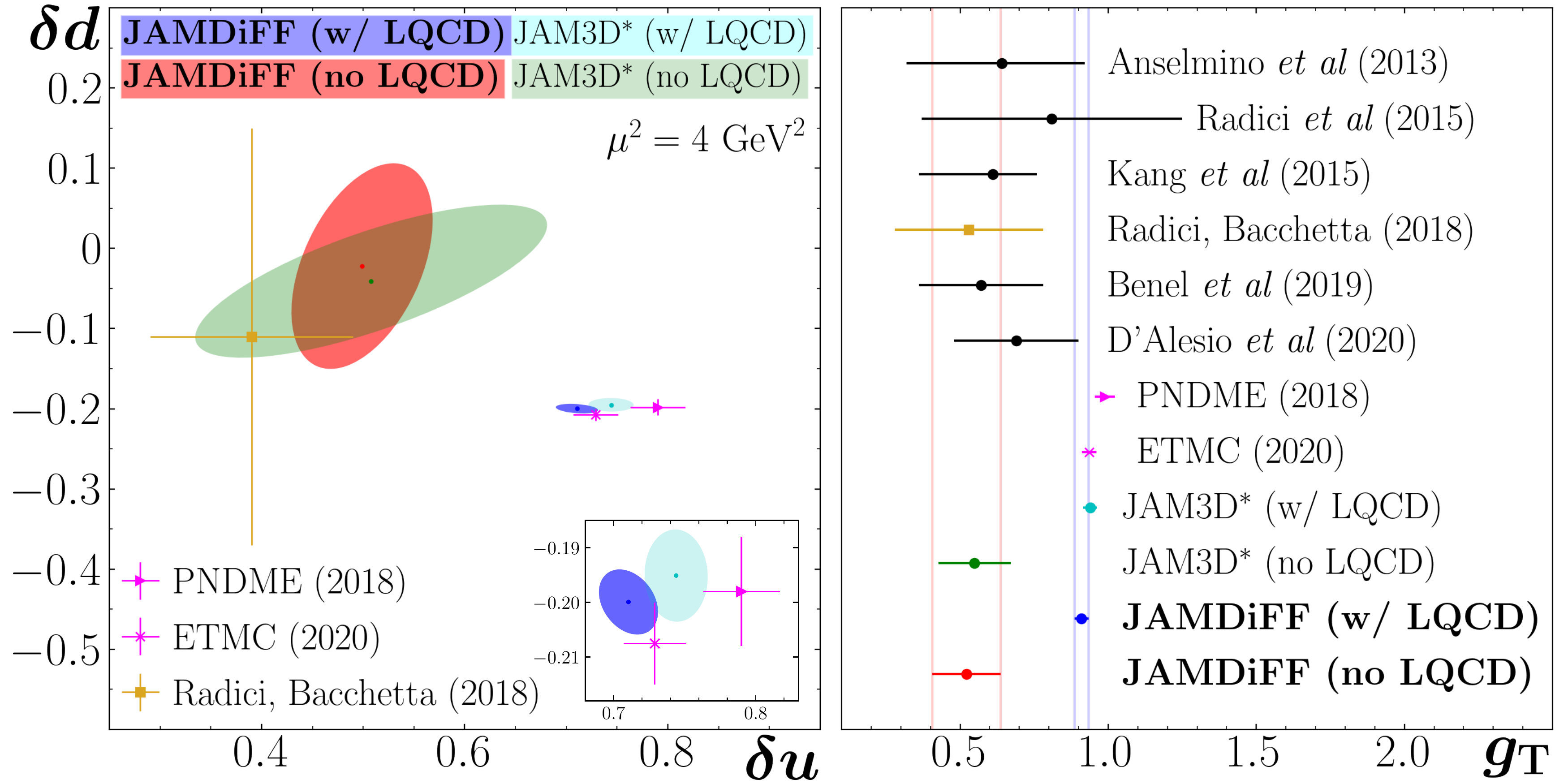}
\caption
{The tensor charges $\delta u$, $\delta d$, and $g_T$.
Our results (JAMDiFF) are shown at 1$\sigma$ with (blue) and without (red) LQCD.
They are compared to the JAM3D$^*$ \cite{Gamberg:2022kdb,Note3} results at 1$\sigma$ with (cyan) and without (green) LQCD, the result of RB18~\cite{Radici:2018iag} (gold square, 90\% confidence level), LQCD computations~\cite{Gupta:2018qil, Gupta:2018lvp, Alexandrou:2019brg}
(magenta points), and other phenomenological extractions 
\cite{Anselmino:2013vqa, Radici:2015mwa, Kang:2015msa, Benel:2019mcq, DAlesio:2020vtw}
(black circles).  The inset shows a close up of the LQCD data and the results from the JAMDiFF and JAM3D$^*$ (both with LQCD)~fits.
All results are at the scale $\mu^2 = 4 \, \textrm{GeV}^2$, except for Anselmino {\it et al} ($2.4 \, \textrm{GeV}^2$), Kang {\it et al} ($10 \, \textrm{GeV}^2$), and Benel {\it et al} ($5 \, \textrm{GeV}^2$).}
\label{f.tensorcharge}
\end{figure}

{\it Tensor Charges ---}\
%
In Fig.~\ref{f.tensorcharge} we show the tensor charges extracted without and with LQCD and compare to other phenomenological analyses and LQCD calculations.
Note that, as discussed above, we use theoretical constraints that limit the PDFs at small $x \lesssim 0.005$ (Eq.~(\ref{e.alpha})) and high $x \gtrsim 0.5$ (the Soffer bound) so that our results for the full moments are not subject to uncontrolled extrapolation errors.
Without LQCD, we find that JAMDiFF, JAM3D$^*$, and RB18 all agree within errors, with our analysis and JAM3D$^*$ preferring a larger $\delta u$ to the RB18 value.
Comparing to LQCD for $\delta u$, we find a 3.2$\sigma$ (3.9$\sigma$) discrepancy with ETMC (PNDME),
while for $\delta d$ we find 1.4$\sigma$ (1.4$\sigma$).
For $g_T$ we find agreement with all other phenomenological results due to large error bars on most extractions, but a 3.5$\sigma$ (3.9$\sigma$) discrepancy with ETMC (PNDME).

For the results with LQCD included in the fit, shown in the inset of Fig.~\ref{f.tensorcharge}, our analysis has no issue in accommodating the lattice result for $\delta d$ (0.8$\sigma$ difference with ETMC and 0.2$\sigma$ with PNDME).
Our result for $\delta u$ agrees with that of ETMC (0.6$\sigma$ difference), but remains smaller than the PNDME data point (2.3$\sigma$ difference).
For $g_T$ (which is not directly included in this analysis) we find a result that is in agreement with ETMC (0.8$\sigma$ difference) but again smaller than PNDME (1.9$\sigma$ difference).
Our tensor charges are summarized in Table~\ref{t.tensorcharges}.

Although our no LQCD result for $\delta u$ is much smaller than the values from ETMC and PNDME, we find that the fit is able to accommodate that lattice data without deteriorating in its description of the experimental measurements.
The noticeable ($\sim\! 3\sigma$) shift in $\delta u$ when including LQCD in the analysis seems surprising at first.
However, while the experimental data has a preference for the size of $h_1^{u_v}$  at large $x \gtrsim 0.3$, this is a mild preference that is easily changed by the inclusion of the LQCD data, as seen in Fig.~\ref{f.tpdfs}.
If additional (precise) experimental data were available at large $x$, it would provide further insight on the behavior of $h_1^{u_v}$ in that region.

Clearly the inclusion of the precise LQCD $\delta u$ data creates a preference for a larger $h_1^{u_v}$ (along with the 2015 STAR data, as demonstrated by its $\chired$ improving with the inclusion of LQCD) than the experimental data alone.
In such a situation where there are competing preferences, and we compare analyses containing different subsets of the data, the choice of likelihood function $\mathcal{L}$ and prior $\pi$ do not guarantee that the fits overlap within statistical uncertainties (see, e.g., Ref.~\cite{Sivia1996DealingWD}).
Before drawing a conclusion about the compatibility between LQCD tensor charges and experimental data, one needs first to include both in the analysis.
One should only be concerned if the description of the lattice data remains poor even after its inclusion and/or if the description of the experimental data suffers significantly.
We refer the reader to Ref.~\cite{Cocuzza:2023vqs} for a more extended discussion about the compatibility between LQCD and experimental measurements based on our results.

We stress that both the present analysis and JAM3D$^*$ find independently that the LQCD results are consistent with the experimental data.
Furthermore, as is seen from Fig.~\ref{f.tpdfs}, the $x$-dependence of the transversity PDFs between this analysis and JAM3D$^*$ are in reasonable agreement.
Overall, this shows that both the DiFF and TMD/collinear twist-3 phenomenological approaches are compatible with LQCD, the Soffer bound, and the small-$x$ asymptotics of transversity, and thus that there is a universal nature of all available information for the transversity PDFs and the tensor charges of the nucleon.

\begin{table}[t]
\resizebox{0.30\textwidth}{!} {
\begin{tabular}{l | c | c | c }
\hhline{====}
\vspace{-0.3cm} & & &  \\
Fit & $\delta u$ & $\delta d$ & $g_T$ \\
\vspace{-0.3cm} & & &  \\
\hline
w/ LQCD & 0.71(2) & -0.200(6) & 0.91(2) \\
no LQCD & 0.50(7) &  -0.02(13) & 0.52(12) \\
\hhline{====}
\end{tabular}}
\caption
{Tensor charges with 1$\sigma$ errors at $\mu^2 = 4$ GeV$^2$.}
\label{t.tensorcharges}
\end{table}

{\it Conclusions ---}\
%
We have presented results of the first QCD global analysis of $e^+e^-$ annihilation, SIDIS, and $pp$ dihadron measurements where both DiFFs and transversity PDFs are extracted simultaneously. 
For the first time, we have studied the Belle cross section~\cite{Belle:2017rwm} (utilizing an improved parameterization for the $D_1$ DiFF), the latest measurements from STAR~\cite{STAR:2017wsi}, and all kinematic variable binnings for the relevant processes under consideration.
We have incorporated theory constraints on transversity at small $x$ and large $x$.
Upon including the LQCD results for $\delta u$ and $\delta d$ from ETMC~\cite{Alexandrou:2019brg} and PNDME~\cite{Gupta:2018lvp}, we find compatibility with this data while maintaining a very good description of the experimental measurements.
Furthermore, our results match those from the single-hadron TMD/collinear twist-3 analysis of JAM3D.  
(In the future, when the theoretical calculations are available, an NLO analysis is needed to definitively confirm these findings.)
We have thus demonstrated, for the first time, the universal nature of all available information on the transversity PDFs and tensor charges of the nucleon.


\begin{acknowledgments}
{\it Acknowledgments} ---
We thank Wally Melnitchouk for helpful discussions, 
Yuri Kovchegov for useful exchanges on the small-$x$ behavior of transversity,
and Anna Martin and Gunar Schnell for clarification on the COMPASS and HERMES data, respectively.
This work was supported by the National Science Foundation under Grants No.~PHY-2110472 (C.C.~and A.M.), No.~PHY-2011763 and No.~PHY-2308567 (D.P.), and No.~PHY-2012002, No.~PHY-2310031, No.~PHY-2335114 (A.P.), and the U.S. Department of Energy contract No.~DE-AC05-06OR23177, under which Jefferson Science Associates, LLC operates Jefferson Lab (A.P.~and N.S.). 
The work of N.S. was supported by the DOE, Office of Science, Office of Nuclear Physics in the Early Career Program.
The work of C.C., A.M., and A.P.~was supported by the U.S. Department of Energy, Office of Science, Office of Nuclear Physics, within the framework of the TMD Topical Collaboration, and by Temple University (C.C. and A.P.).
\end{acknowledgments}


\bibliography{cc.bib}{}
\end{document}